\documentclass[review,12pt,3p]{elsarticle}
\usepackage{hyphenat}

\makeatletter
\def\ps@pprintTitle{%
	\let\@oddhead\@empty
	\let\@evenhead\@empty
	\def\@oddfoot{}%
	\let\@evenfoot\@oddfoot}

\usepackage{multirow}
\usepackage[
singlelinecheck=false 
]{caption}
\usepackage{enumitem}

\usepackage{enumitem}
\usepackage{bm}
\usepackage{setspace}
\usepackage{makecell}

\usepackage{amsmath}

\usepackage{amssymb}

\begin{document}
	
	\begin{frontmatter}
		\title{On the monotonicity of the critical time in the Constrained-degree percolation model  \\} 

		\author[label1]{Charles S. do Amaral\corref{cor1}}
		\address[label1]{Departamento de Matem\'atica - Centro Federal de Educa\c c\~ao Tecnol\'ogica de Minas Gerais, CEFET--MG, Av. Amazonas 7675, Belo Horizonte, MG, Brasil. }
		\ead{charlesmat@cefetmg.br}
		\cortext[cor1]{Corresponding author.}
		
		\author[label7]{A. P. F. Atman}
		\address[label7]{Departamento de F\'isica - Centro Federal de Educação Tecnológica de Minas Gerais, CEFET-MG and Instituto Nacional de Ci\^encia e Tecnologia - Sistemas Complexos.}

		\author[label6]{Bernardo N. B. de Lima}
		\address[label6]{Departamento de Matem\'atica, Universidade Federal de Minas Gerais, UFMG, Av. Ant\^onio Carlos, 6627, Belo Horizonte, MG, Brasil.}

		\begin{abstract}
			The Constrained-degree percolation model was introduced in [B.N.B. de Lima, R. Sanchis, D.C. dos Santos, V. Sidoravicius, and R. Teodoro, Stoch. Process. Appl. (2020)], where it was proven that this model has a non-trivial phase transition on a square lattice. We study the Constrained-degree percolation model on the $d$-dimensional hypercubic lattice ($\mathbb {Z}^d$) and, via numerical simulations, found evidence that the critical time $ t_{c}^{d}(k) $ is monotonous not increasing in the constrained $k$ if $d=3,4$, like it is when $d=2$. We verify that the lowest constrained value $k$ such that the system exhibits a phase transition is $k=3$ and that the correlation critical exponent $\nu$ for the Constrained-degree percolation model and ordinary Bernoulli percolation are the same. 
		\end{abstract}

		
	\end{frontmatter}
	
	\section{Introduction}

	The classical percolation model was proposed by Broadbent and Hammersley in 1957~\cite{broadbent} with the idea of modeling the flow of a fluid (deterministic) through a porous medium (random environment). In the ordinary Bernoulli bond percolation model, with parameter $p$, on a graph, each bond is  \textit{open} or \textit{closed}, independently, with probabilities $p$ and $1-p$, respectively. For values of $ p $ greater than one constant $p_c$, called \textit{critical point} (percolation threshold), an infinite sequence of connected open bonds (sites) appears. In this case, we say that \textit{percolation} has occurred. There are several variations of these models in the literature with applications in different areas of science, in [2-17] there are some examples.
	
	The present study aims to investigate properties of one of these variations, the so called Constrained-degree bond percolation model (CDPM) introduced in \cite{de_lima}. The CDPM is defined in a infinite graph $\mathcal{G} = (\mathcal{V}, \mathcal{E})$ connected and with bounded degree. The number of bonds that has some end on the vertex $ v $ is denoted for $\mbox{deg}(v)$ (degree of $v$). We fix a sequence of parameters (capacities or constrains) $(k_v)_{v \in \mathcal{V}}$, with $k_v \leq \mbox{deg}(v)$, and consider a time continuous percolation processes defined as follows: let $(U_e)_{e\in\mathcal{E}}$ be a sequence of independent and identically distributed random variables with uniform distribution on $[0,1]$. At the time $t=0$ all bonds are closed. Each bond $ e=<u,v> $ will try to open at time $U_e$, it will succeed if and only if $\overline{\mbox{deg}}(u,U_e) < k_{u}$ and $\overline{\mbox{deg}}(v, U_e) < k_{v}$, where $\overline{\mbox{deg}}(v, t)$ denotes the degree of $v$ considering only the open bonds at time $t$. We say that a site $ v $ percolates in time $ t $ if there is an infinite path of open bonds starting from the vertex $v$. Observe that when $k_v=\mbox{deg}(v),\ \forall v\in\mathcal{V}$, the CDPM reduces to ordinary Bernoulli bond percolation; otherwise, it is a dependent model whose dependence has infinite range, in particular the Fortuin-Kasteleyn-Ginibre (FKG) inequality (see \cite{grimmett}) does not work.

	In 1979, Gaunt proposed \cite{gaunt} a model called \textit{Percolation with restricted-valence} which is analogous to the CDPM, but considering site percolation instead of bond percolation. Kertész, Chakrabarti and Duarte \cite{kertesz} estimated the values of critical points and the \textit{maximum random concentration} for this model on the hypercubic lattice $ \mathbb{Z}^d $ for $ d=2,3 $. The CDPM is related to the study of dimers \cite{furlan} and polymers \cite{soteros, wilkinson1}.  Other types of constrained percolation models, where only specific configurations are allowed on the vertices, were studied in \cite{holroyd, garet}. 
	
\begin{figure}[t]
		\begin{center}
			\includegraphics[height=7.0cm]{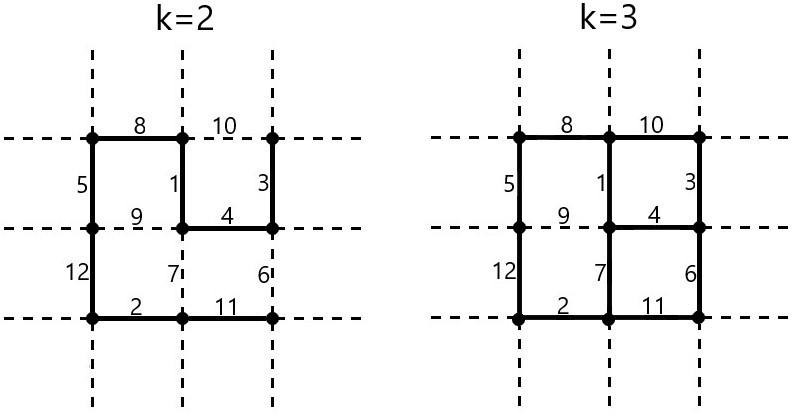}
			\caption{Part of the square lattice $\mathbb{Z}^2$ in which the order of the twelve first bonds that we will try to open is indicated. Bonds represented with dotted lines are closed and continuous lines are open. The configuration shown occurs when the twelfth bond is opened.}
			\label{model_example}
		\end{center}
	\end{figure}

	From now on, we consider the CDPM on the hypercubic lattice $\mathbb{Z}^d$ and $k_v=k,\ \forall v\in\mathbb{Z}^d$. In Figure \ref{model_example} we show a part of $ \mathbb{Z}^2 $ considering the same sequence $ (U_e)_{e \in \mathcal{E}} $ for the cases $ k = 2,3 $. 
	
	Let us denote by $\theta^d(t;k)$ the probability of the origin of $\mathbb{Z}^d$ percolates at time $t$ with constrained $k$. Since the function $\theta^d(t;k)$ is monotone in $t$, we can define the critical time as:
	\begin{equation}
	t_{c}^d(k) = \sup\{ t \in [0,1]; \theta^d(t;k)=0 \}.
	\label{def_tc}
	\end{equation}
	
	\noindent If $\theta^d(1;k)=0$, there is no percolation for any value of $ t $ and we define $ t_{c}^d(k) = \infty $. We denote by $\psi^{d}(t;k)$ the probability of there is at least one vertex that percolates. It is standard to show that $\theta^{d}(t;k)>0$ if and only if $\psi^{d}(t;k)=1$, since the underlying probability measure is translation invariant, that is $\psi^{d}(t;k)\in\{0,1\}$. So we can replace the function $\theta^{d}(t;k)$ in (1) by $\psi^{d}(t;k)$.
	
	As already mention, when $ k=2d $ this model is the ordinary Bernoulli bond percolation model with parameter $t$, therefore $ t_c^d(2d) = p_c (\mathbb{Z}^d)$ (the ordinary percolation threshold). For the case $k=1 $ there will be only isolated bonds, thus percolation does not occur for all $ t \in [0,1]$ and $d \geq 1$. It was proven in \cite{de_lima} that if $k=2$, $t_{c}^{d}(2)=\infty$ for all $d \geq 2$, that is, percolation does not occur for all times, included $ t = 1 $. The main result in \cite{de_lima} states that for the square lattice ($d=2$) $t_c^2(3)\in(\frac{1}{2},1)$, characterizing a phase transition for $k=3$. Apart these results, there are many questions and few answers.
	
\begin{figure}
		\begin{center}
			\includegraphics[height=8.0cm]{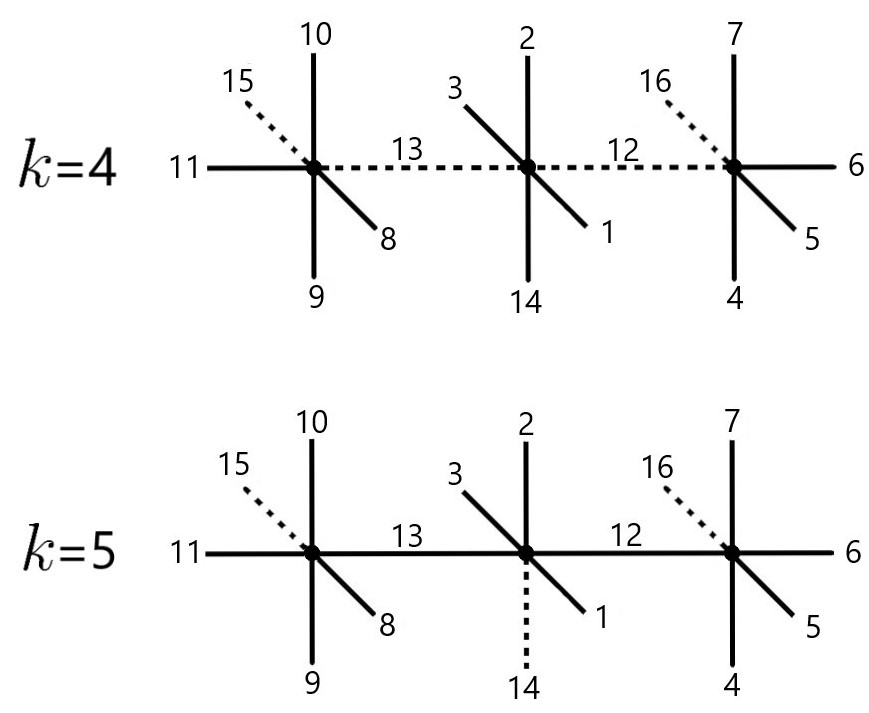}
			\caption{A piece of $\mathbb{Z}^3$ where the numbers represent the order in which the bonds will be try to open. When $t$ is equal to the opening time of the fourteenth bond, we have that it will open if the constraint is $ k = 4 $, but the same bond will be closed if $ k = 5 $.}
			\label{example2}
		\end{center}
	\end{figure}

	One of these intriguing facts is the following: fixed the dimension $d$, is the critical time a non-increasing function in $k$? that is, $t_c^d(k+1)\leq t_c^d(k)$? There is no trivial coupling proving this inequality (in Figure \ref{example2} we present an example of this situation). If the sequence $(t_c^d(k))_k$ is indeed monotone in $k$, we could define, for each dimension $d>2$, the critical restriction $k_c^d=\inf\{k;t_c^d(k)\leq 1\}$. The main purpose of this work is to perform simulations of $t_c^d(k)$ for several values of $d$ and $k$ that support the  monotonicity of the sequence $(t_c^d(k))_k$ as well as $k_c^d=3,\ \forall d\geq 2$. Observe that this is true for the square lattice, as shown in \cite{de_lima}, it holds that $t_c^2(1)=t_c^2(2)=\infty,\ \frac{1}{2}<t_c^2(3)<1$ and $t_c^2(4)=\frac{1}{2}$ (the ordinary percolation threshold for the square lattice proved by Harris \cite{harris} and Kesten \cite{kesten}). 
	
	Using an algorithm based on the Newmann-Ziff algorithm~\cite{ziff01}, we determine $t_{c}^{d}(k)$ for $k~\in~\{2,3..., 2d\}$ ($d=2,3,4$) and the maximum random concentration ($ x_k $), which is defined as the expected value of open bonds in the lattice when $ t = 1 $. In \cite{gaunt}, $x_k$ is the expected value of open sites instead of bonds.

	As mentioned before, in the CDPM, when $k<2d $, the probability of a bond being open or closed depends on the status of the other bonds (long-range dependence), differently from what occurs in the ordinary Bernoulli percolation model. Therefore, it is not possible to use algorithms that assume that each bond is open (closed) independently, for example, the Leath Algorithm \cite{leath} or Invasion Percolation Algorithm \cite{lenormand, chandler, wilkinson2}. Such algorithms generate only a single cluster, rather than an entire lattice configuration and are often used to estimate the critical point of percolation models in higher dimensions, as they have low time and space computational complexities.

	We find numerical evidence that when $ d = 3,4 $, $ t_{c}^d (k) $ is monotonous in $ k $ and we determine that $ k_{c}^d=3$ for these cases. In addition, we also simulated the case $d=2$, verifying the results of \cite{de_lima} and obtaining numerically the value of $t_c^2(3)$ (that was also estimated in \cite{furlan}). Based on this results, we conjecture that $k_{c}^{d}=3$ and $t_{c}^{d}(k)$ is monotonous in $k$ to every $ d \geq 2 $. Furthermore, through the analysis of \textit{correlation length exponent} $\nu$, we found evidence that probably the CDPM (for $k \geq 3$) and the classic percolation model are in the same universality class. In \cite{furlan} it is shown that this occurs, when $ d = 2 $, through a detailed analysis of the critical exponents.

	The remainder of this paper is organized as follows. Section 2 describes our simulation procedure and Section 3 discusses the results obtained. Conclusions are summarized in Section 4.

	\section{Numerical Procedure}
	
	We consider the regular box $L \times ... \times L$ on $\mathbb{Z}^d$ with periodic boundary. The percolation criterion assumed is that the infinite cluster emerges in the box when there exists a cluster that wraps around one of the $d$ directions. From now on, we will omit $d$ and $k$ in the $t_{c}^{d}(k)$ notation if they are clear from the context.
	
	We denote by $\psi_{L}(t;k)$ the probability of percolation in the box of length $L$. For each configuration $ (U_e)_{e \in \mathcal{E}} $, we define $\mathcal{O}_t$ as the number of bonds that we will try to open until time $t$, formally we have:
	\begin{equation}
	\mathcal{O}_t = \# \{e \in \mathcal{E}; U_e \leq t \}
	\end{equation}
	
	\noindent where the $\#$ symbol denotes the number of elements in the set. Note that the probability of a bond belongs to $\mathcal{O}_t$ does not depend on the state of any other bond (the same does not occur when we consider the set formed only by bonds opened up to time $t$). This independence allows us to write the equation (\ref{psi_L}) and estimate $t_{c}$ as follows.

	Using the \textit{Partition Theorem} \cite{grimmett_prob}, the probability of percolation in the box of length $L$ can be write as:
	\begin{equation}
	\psi_{L}(t; k)= \sum_{i=0}^{N} \mathbb{P} (\mathcal{O}_t=i) \cdot \overline{Q_i} = \sum_{i=0}^{N} 
	\left(\begin{array}{c}
	N \\
	i 
	\end{array}\right) \cdot t^{i} \cdot (1-t)^{(N-i)} \cdot \overline{Q_i}
	\label{psi_L}
	\end{equation}
	
	\noindent where $\overline{Q_i}$ is the probability of percolation, at time $t$, conditioned that $\mathcal{O}_t=i$. Differentiating (\ref{psi_L}) \cite{choi}, it holds that
	\begin{equation}
	\dfrac{d\psi_{L}(t; k)}{dt}= \sum_{i=0}^{N} 
	\left(\begin{array}{c}
	N \\
	i 
	\end{array}\right) \cdot (i-Nt)\cdot t^{i-1} \cdot (1-t)^{(N-i-1)} \cdot \overline{Q_i}.
	\label{dpsi_L}
	\end{equation}
	
	To estimate $\overline{Q}_{i}$, we perform an analysis similar to that used in the Newmann-Ziff algorithm~\cite{ziff01}. A single run of the algorithm is obtained as follows:

	\begin{figure}[t]
		\begin{center}
			\includegraphics[height=6.8cm, width=16.5cm]{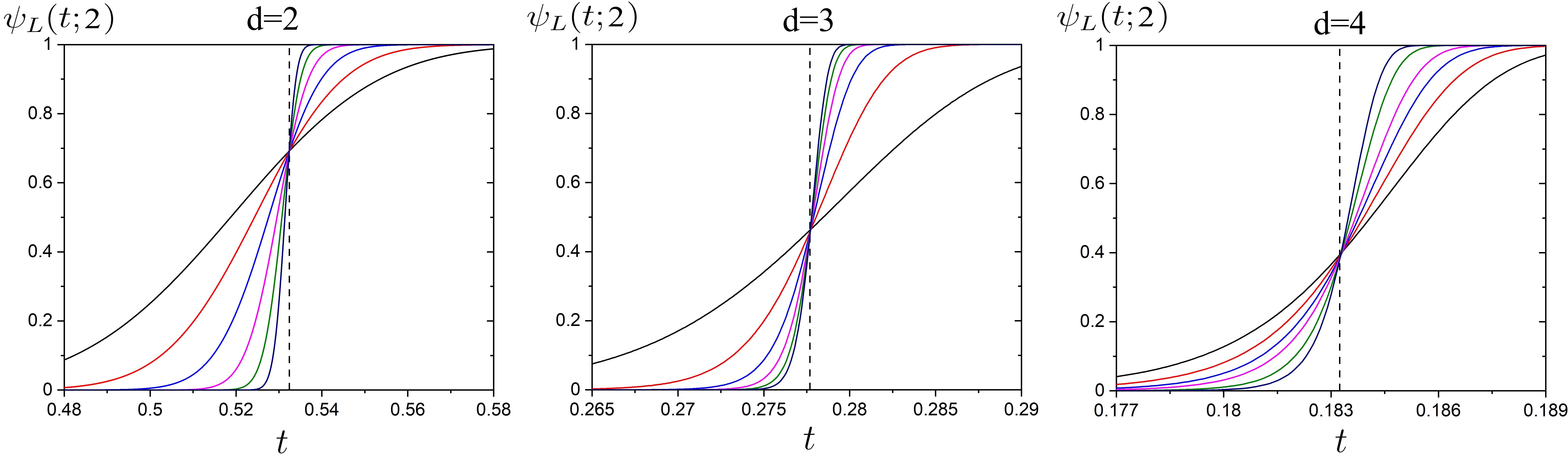}
			\caption{Graphs of the $\psi_{L}(t; k)$ functions when $k=3$ and $d=2,3,4$. The dashed line marks the critical time obtained using the FSS (\ref{tc}). The graph's slope near the critical time grows as the $ L $ increases. }
			\label{psi}
		\end{center}
	\end{figure}
	
	\begin{itemize}
		\item[(i)]  Create a list with a permutation of all the $ N = dL^d $ bonds ($ e_1, e_2, ..., e_N $).
		\item[(ii)] Try to open the bonds following the order $ e_1 $, ..., $ e_N $ and check if percolation has occurred. If a bond does not satisfy the $ k $ constraint, then it remains closed.  
		\item[(iii)] If percolation occurred when the bond $ e_i $ was opened, then we define $Q_j = 1$ for $j \geq i$. Otherwise $ Q_i = 0 $.
	\end{itemize}
	
	\noindent The estimate of $\overline{Q}_{i}$ is obtained as the mean of a large number of samples of $Q_i$.
	
	We study system with sizes $32 \leq L \leq 1024$ ($d=2$), $22 \leq L \leq 176$ ($d=3$) and $18\leq L \leq 46$ ($d=4$), with 6 values of $L$ for each $d$. The number of samples varies between $50000$ ($L=48$, $d=4$) until $10^6$ ($L=32$, $d=2$). 
	
	For each $k$ and $d$, the critical time ($t_c$) is obtained by the finite-size scaling (FSS)  
	\begin{equation}
	|\overline{t}_L-t_c| \sim L^{a},
	\label{tc}
	\end{equation}
	
	\noindent where $\overline{t}_L=\int_{0}^{1} t \cdot \frac{d\psi_{L}(t; k)}{dt} dt$ is the average concentration when percolation occurs for the first time and $a$ is a fitting parameter \cite{stauffer94}. The exponent $\nu$ was estimated from the scaling relation
	\begin{equation}
	\mbox{Max}\left[\dfrac{d\psi_{L}(t; k)}{dt} \right] \sim L^{-\frac{1}{\nu}},
	\label{nu}
	\end{equation}
	
	\begin{table*}[t]
		\centering
		\begin{small}
			\caption*{\ \textbf{Table 1:} Values obtained for $\frac{1}{\nu}$ and $t_c$. When $d=4$, the central values obtained for $t_c(7)$ and $t_c(8)$, marked with $^{*}$, were $ 0.16011996$ and $0.16011964$, respectively.}
			\label{tabela1}
			\begin{tabular}{|c|cc||c|cc||c|cc|}
				\hline
				\multicolumn{3}{|c||}{$d=2$} & \multicolumn{3}{c||}{$d=3$} & \multicolumn{3}{c|}{$d=4$} \\
				\hline
				$k$    &      $t_c$ &   $\frac{1}{\nu}$  &  $k$  &    $t_c$ &  $\frac{1}{\nu}$ &  $k$  &   $t_c$ &  $\frac{1}{\nu}$  \\  \hline
				\multirow{4}{*}{3}      &        \multirow{4}{*}{0.532393(15)}   &        \multirow{4}{*}{0.749(4)}    &  \multirow{1.5}{*}{3}    & \multirow{1.2}{*}{0.277691(10)}   &   \multirow{1.2}{*}{1.15(2)}    &  \multirow{1.2}{*}{3}  & \multirow{1.2}{*}{0.183205(30)} & \multirow{1.2}{*}{1.47(5)} \\
				\multirow{7}{*}{4}  &  \multirow{7}{*}{0.499994(16)}   &     \multirow{7}{*}{0.751(3)}      &   \multirow{3}{*}{4}   & \multirow{3}{*}{0.251319(6)} & \multirow{3}{*}{1.15(2)}     & \multirow{1.4}{*}{4}  &    \multirow{1.4}{*}{0.162716(22)}    &     \multirow{1.4}{*}{1.46(5)} \\  
				&    &         &  \multirow{4.7}{*}{5}  & \multirow{4.7}{*}{0.248922(8)} & \multirow{4.7}{*}{1.14(2)}    & \multirow{1.8}{*}{5}    &  \multirow{1.8}{*}{0.160348(15)}    &  \multirow{1.8}{*}{1.47(3)}     \\
				&   &       &      &  &     & \multirow{2.3}{*}{6}     &  \multirow{2.3}{*}{0.160131(14)}  &  \multirow{2.3}{*}{1.47(4)}  \\
				&   &       &  \multirow{4.8}{*}{6}      & \multirow{4.8}{*}{0.248810(7)} &  \multirow{4.8}{*}{1.14(2)}   & \multirow{2.8}{*}{7}    &   \multirow{2.8}{*}{\ 0.160120$(15)^{*}$}    &  \multirow{2.8}{*}{1.46(4)}  \\
				&   &      &        & &     & \multirow{3.2}{*}{8}    &  \multirow{3.2}{*}{\ 0.160120$(15)^{*}$}    &   \multirow{3.2}{*}{1.47(4)}   \\
				&   &      &       &  &  &       &  & \\
				\hline
			\end{tabular}
		\end{small}
	\end{table*}
	
	\noindent where the left-hand side represents the maximum of $\frac{d\psi_{L}(t; k)}{dt}$ for size $L$.

	We obtained five independent estimates for each parameter studied; the central value is the average of them. The uncertainties were obtained using the standard deviation of the mean.

	\section{Results}
	
	Although analytical or numerical results already exist, we also simulate the cases $k=2$ and $k=2d$ for each $d$. When $k=2$ we found that $t_c=\infty$ for all $d$, as obtained analytically in \cite{de_lima}. When $d=2$ and $k=4$ , we get $t_c=0.499994(16)$ that is consistent with the exact value $t_c=\frac{1}{2}$ (see \cite{grimmett} for a rigorous proof). For $d=3$ and $k=6$, we estimated $t_c=0.248810(7)$; for $d=4$ and $k=8$, we estimated $t_c=0.160120(15)$. These values are consistent with the numerical results for the ordinary Bernoulli percolation threshold obtained for $d=3$: $0.2488126(5)$ \cite{lorenz_d3_1}, $0.2488125(25)$ \cite{dammer_d3_2} and $0.24881182(10)$ \cite{wang_d3_3}; and for $d=4$: $0.1601310(10)$ \cite{dammer_d3_2}, $0.16013122(6)$ \cite{mertens_d4_1} and $0.1601312(2)$ \cite{xun_d4_2}. 
	
	\begin{figure}[t!]
		\begin{center}
			\hspace{0.11cm}\includegraphics[height=5cm, width=8.5cm]{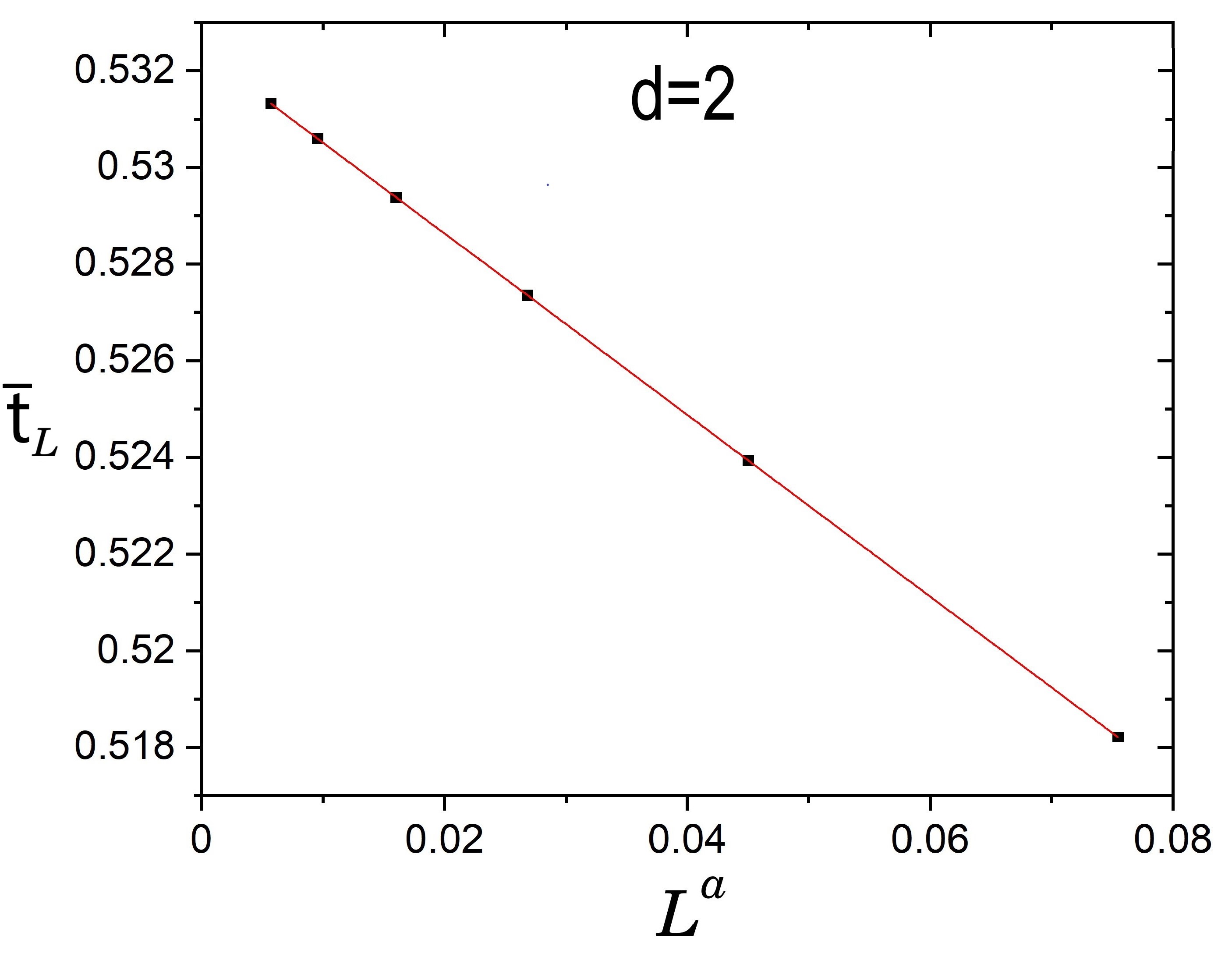} \quad
			\includegraphics[height=5cm, width=7.3cm]{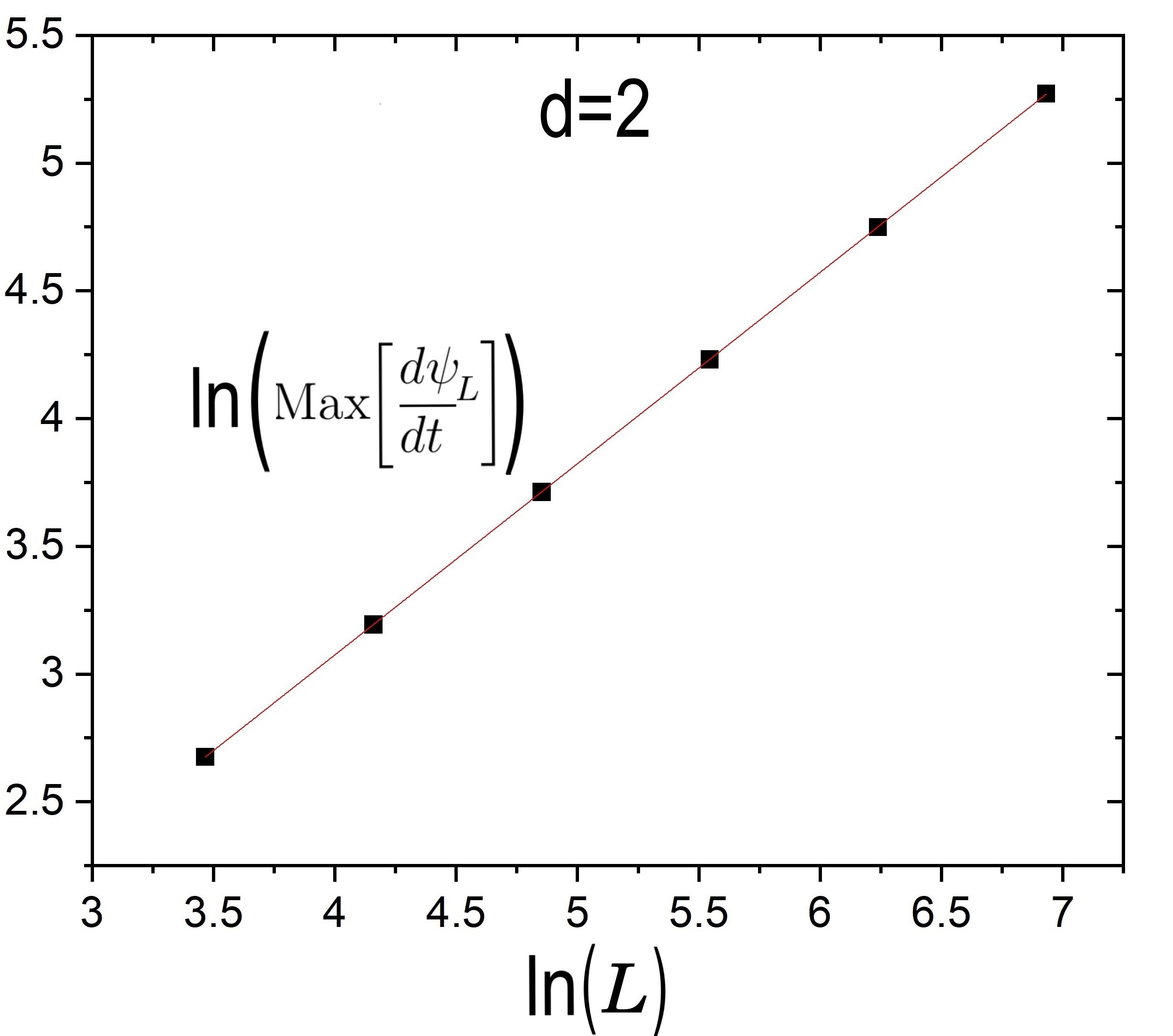} \\[10pt]
			\includegraphics[height=5cm, width=8.5cm]{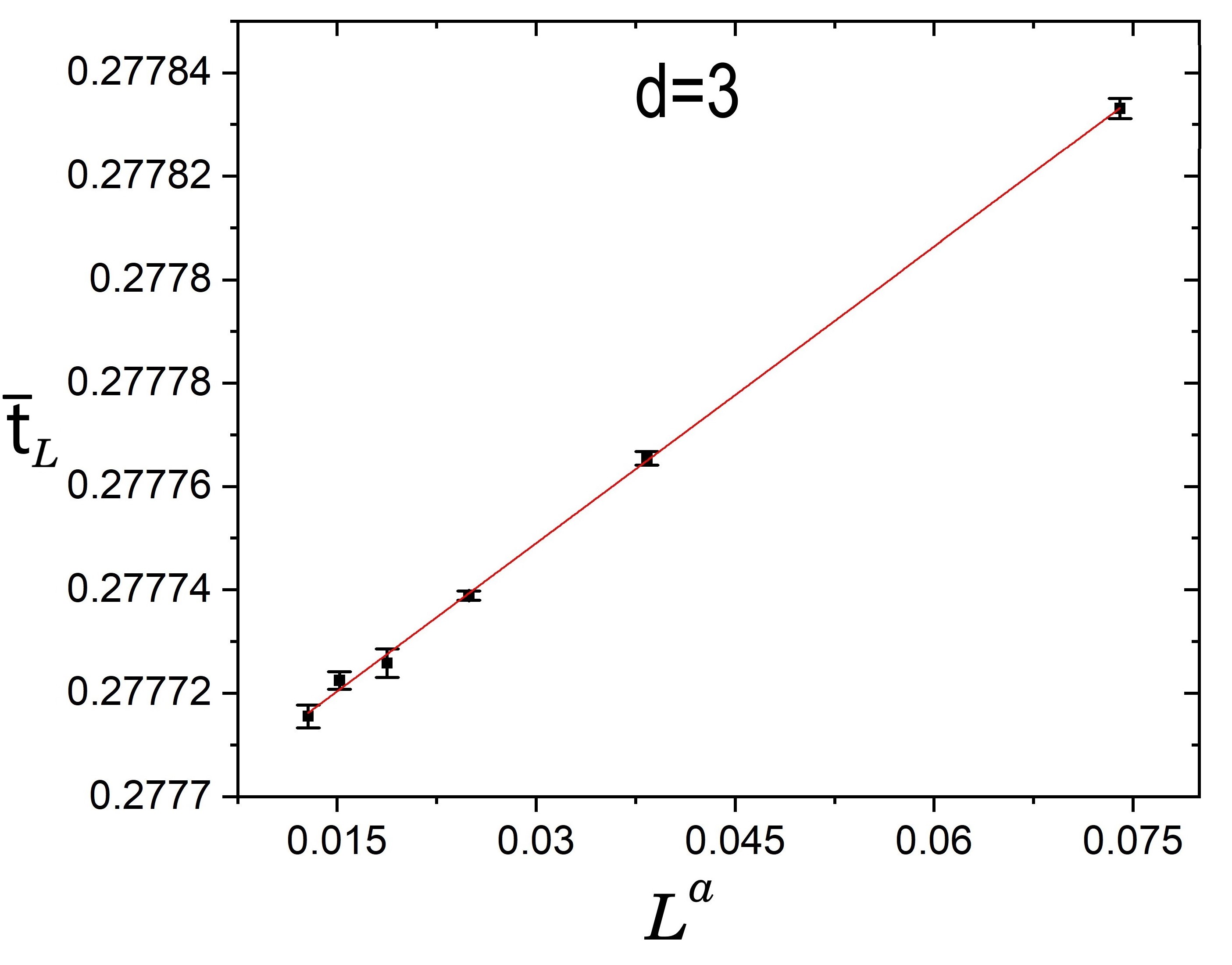} \quad
			\includegraphics[height=5cm, width=7.3cm]{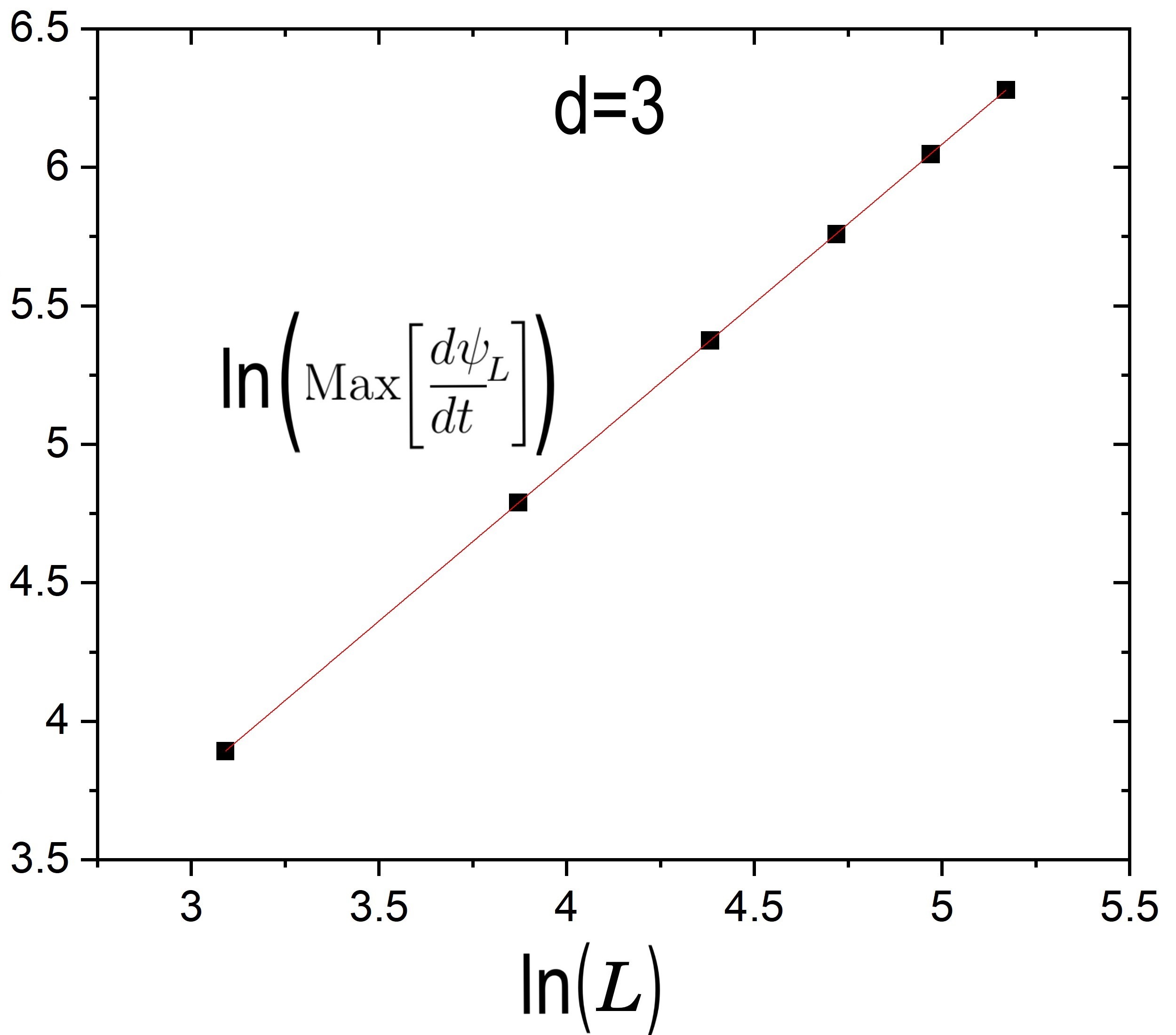} \\[10pt]		
			\includegraphics[height=5cm, width=8.5cm]{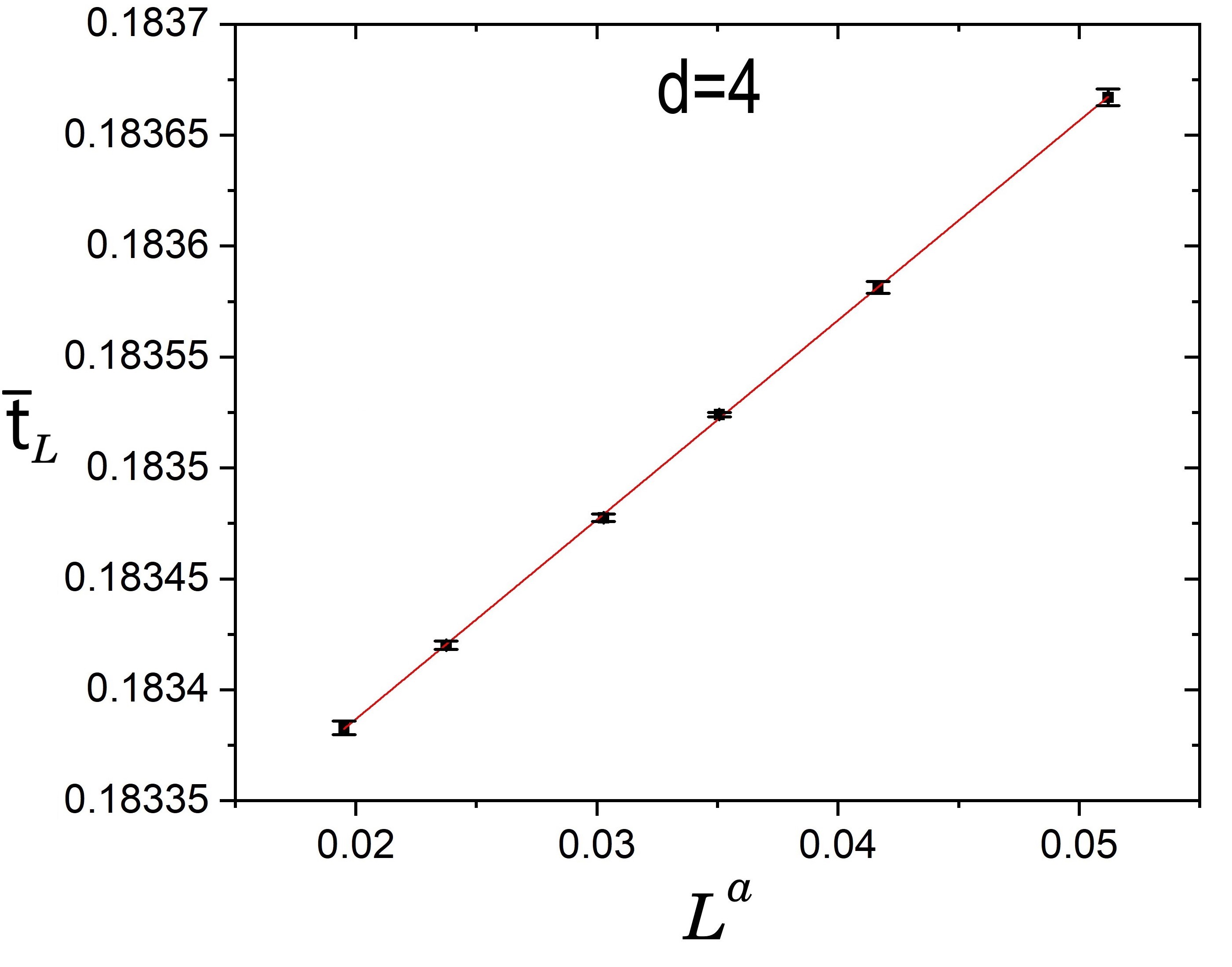} \quad
			\includegraphics[height=5cm, width=7.3cm]{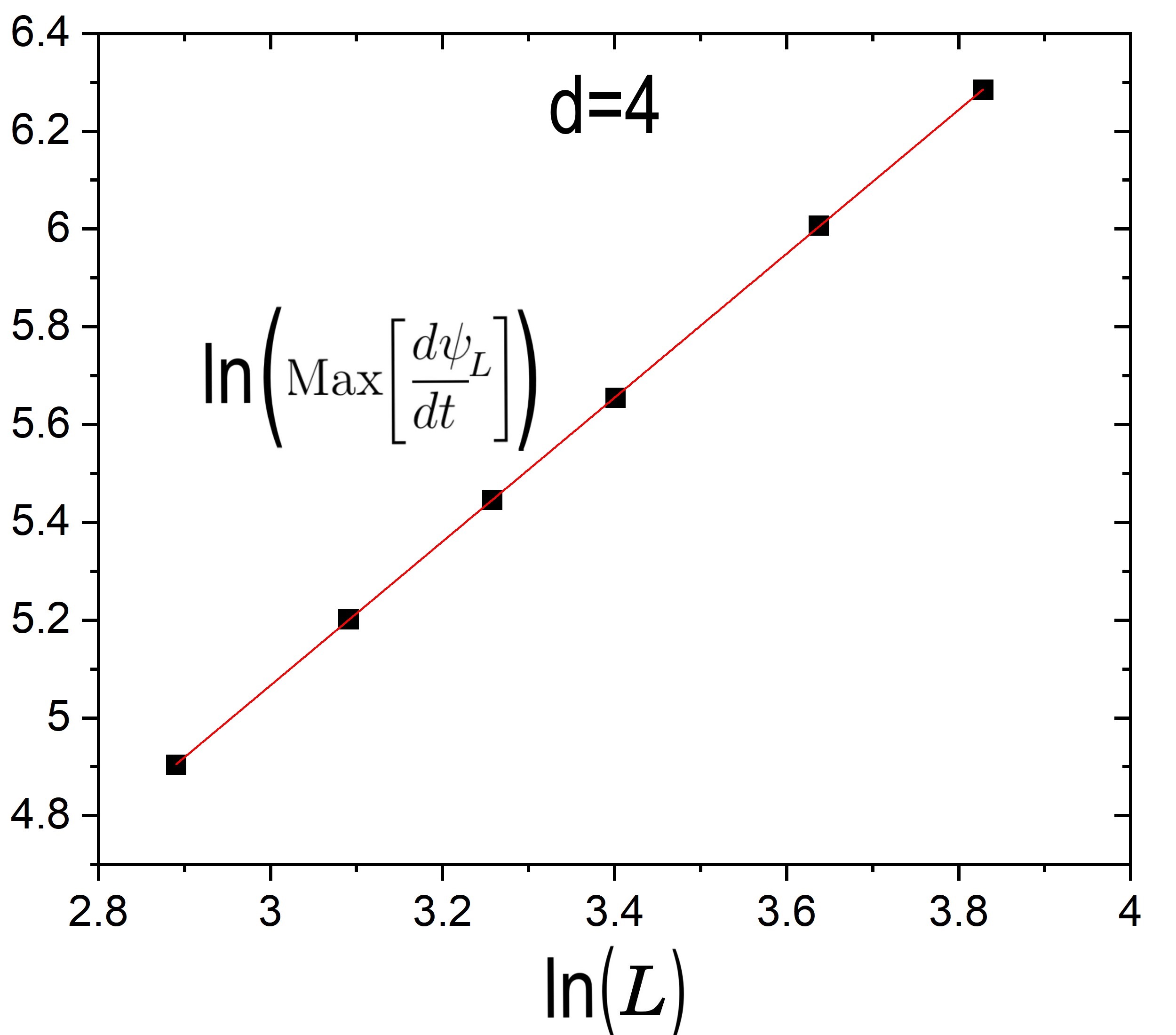} 
			\caption{Graphs obtained for $k=3$, considering $d=2,3,4$, to determine $t_c$ and $\frac{1}{\nu}$. The left plots were obtained using the FSS (\ref{tc}), and the right plots through log-log plots using (\ref{nu}). The error bars are smaller than the symbols in the graphics where they are not shown.}
			\label{k3}
		\end{center}
	\end{figure}

	For the case $k=3$ when $d=2,3,4$, Figure \ref{psi} presents graphs of the function $\psi_{L}(t,k)$, and Figure \ref{k3} the graphs obtained through the relations (\ref{tc}) and (\ref{nu}) to estimate $\frac{1}{\nu}$ and $t_c$, respectively. Table 1 shows the results obtained for $\frac{1}{\nu}$ and $t_c$ for all $k$ and $d$ values analyzed. For $d=3$ it becomes clear that $t_c$ is monotonous in $k\geq 3$. If $ d = 4 $, the central values also suggest such behavior, however, considering the uncertainties obtained, we can only say that such behavior will occur for $k \leq 6$. 
	
	In order to obtain more evidence of the possible monotonicity from $t_c(k)$ when $k \geq 6 $, we simulated the model, for $ L = 46 $, considering $M=2 \times 10^5$ configurations $U^{i}=\{U^{i}_{e}\}_{e \in \mathcal{E}}$, $1 \leq i \leq M$. Once a configuration $U^{i}$ is fixed, we denote by $t(k;i)$ as the time that the system percolated for constraint $k$. We found that $t(8;i) \leq t(7;i) \leq t(6; i)$ for all $1\leq i\leq M$. Furthermore, $t(8;i)<t(7;i)$, $t(7;i)<t(6;i)$ and $t(8;i) < t(6;i)$ for $0.096\%$, $3.429\%$ and $3.52\%$ of the analyzed configurations, respectively. Such results support the statement that the critical time will also be monotonous, in $3\leq k \leq 8$, when $d=4$. 
	
	The values obtained for the critical exponent $\nu$, for all $k$, are compatible with the values of ordinary Bernoulli percolation model in $d=2$ ($\frac{4}{3}$ \cite{stauffer94}), $d=3$ ($0.87619(12)$ \cite{xu}, $0.8774(13)$ \cite{koza}) and $d=4$ ($0.6845(23)$ \cite{zhang}). This fact suggests that probably the CDPM is in the same universality class of the ordinary Bernoulli percolation model for all $d$ and $ k \geq 3$.

\begin{figure}[t]
		\begin{center}
			\includegraphics[height=7.5cm]{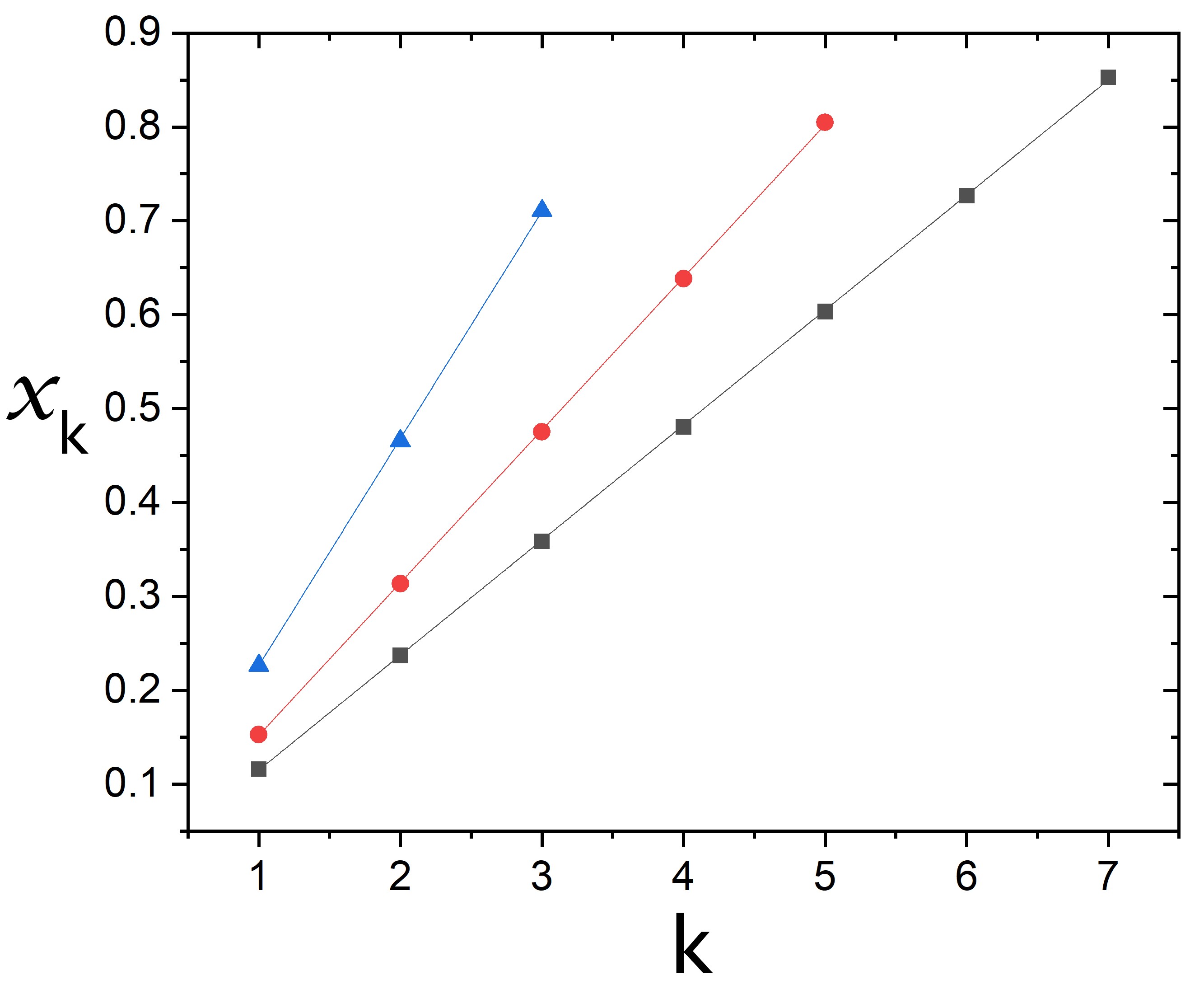} 
			\caption{Estimates for maximum random concentration. Dimension $d=2$, triangles; $d=3$, circles; $d=4$, squares. The slopes obtained were $0.2419(16)$, $0.1626(6)$, and $0.1225(3)$ for $d=2,3,4$, respectively. The error bars are smaller than
				the symbols.}
			\label{lin}
		\end{center}
	\end{figure}

	The estimates obtained for $x_k(L)$, for $k=1,2,3$, when $d=2$ show that its value varies less than $0.009\%$ when we vary $L$ from $32$ to $1024$. In addition, the uncertainty in the estimate decreases from $\approx 3 \times 10^ {-3}$ to $\approx 8 \times 10^{-5}$. For this reason, we estimate $x_k(L)$ only for the highest value of $L$. Table 2 lists the values obtained. Through a plot of $x_{k}$ versus $k$, we see that for all $d$ its values grow linearly (Figure \ref{lin}).

	\begin{table*}[]
		\centering
		\begin{small}
			\caption*{\ \textbf{Table 2:} Estimated random maximum concentration ($x_k$) for all $ k $ and $ d $ values.}
			\begin{tabular}{c|ccccccc|}
				\cline{2-8}
				& $x_1$      & $x_2$      & $x_3$      & $x_4$      & $x_5$      & $x_6$      & $x_7$      \\[2pt]  \hline
				\multicolumn{1}{|c|}{$d=2$} & 0.22670(6) & 0.46578(8) & 0.71094(5) & -          & -          & -          & -          \\[3pt] 
				\multicolumn{1}{|c|}{$d=3$} & 0.15307(2) & 0.31346(2) & 0.47511(3) & 0.63817(3) & 0.80474(3) & -          & -          \\[3pt]  
				\multicolumn{1}{|c|}{$d=4$} & 0.11611(2) & 0.23710(2) & 0.35869(2) & 0.48070(2) & 0.60324(3) & 0.72672(2) & 0.85274(2) \\ \hline
			\end{tabular}
		\end{small}
	\end{table*}

	\section{Conclusion}
	
	We study proprieties of the critical time for the Constrained-degree percolation model on the hypercubic lattice $\mathbb {Z}^d $ when $d=2,3,4$. It has a long-range dependence and applications in the study of dimers and polymers. One interesting problems related to this model is to check if $t_c(k)$ is a monotone function in the constrained parameter $k$; if the answer is affirmative, it is natural also to ask what is the smallest value of $k$ such that the model has a non-trivial phase transition ($t_c \leq 1$). When $k=1$ there will be only isolated bonds (there is no percolation) and when $k = 2d$ this model reduces to ordinary Bernoulli bond percolation, therefore the non-trivial cases occur when $k \in \{2, ..., 2d-1\}$.
	
	In \cite{de_lima}, it was proven that: {\it (i)} for all $d\geq 2$ and $k=2$ there is no percolation or an time $t$ and {\it (ii)} for the squares lattice $d=2$, when $k=3$ it holds that $\frac{1}{2}<t_c<1$, that is, there is a non-trivial phase transition (the exact value of $t_c$ is not known when $ k=3$). 
	
	In this work we determined, via numerical simulations, the value of $t_c$ for all non-trivial cases and observed there is a phase transition for constrained $k=3$, for $d=2,3,4$. In addition, we find numerical evidence that $t_c$ is monotonous in $k$,  and that the expected value of open bonds in the lattice when $t=1$ (maximum random concentration, $x_k$) grows linearly in $k$. We also estimated the correlation length exponent $\nu$ and obtained values compatible with those of the ordinary Bernoulli percolation model, suggesting that both models probably are in the same universality class.
	
	This present study can contribute to the analysis of applied models related to the Constrained-degree percolation model on $\mathbb{Z}^{d}$, besides helping researchers in the analytical and numerical study of this model.

	\section*{Acknowledgments}
	We thank the authors of \cite{de_lima} for the example of Figure 2. The research of A.P.F.A.\ is supported in part by CNPq grant 308792/2018-1 and FAPEMIG. The research of B.N.B.L.\ is supported in part by CNPq grant 305811/2018-5 and FAPERJ (Pronex E-26/010.001269/2016).


\begin{thebibliography}{00}
		
		\bibitem{broadbent}
		S. R. Broadbent and J. M. Hammersley, Proc. Camb. Philos. Soc. \textbf{53}, 629 (1957).
		
		
		
		\bibitem{kirkpatrick}
		S. Kirkpatrick, Rev. Modern Phys. \textbf{45}, 574 (1973).
		
		
		\bibitem{shklovskii}
		B.I. Shklovskii and A.L. Efros, Usp. Fiz. Nauk \textbf{117}, 401(1975).
		
		
		\bibitem{dotsenko}
		V.S. Dotsenko, P. Windey, G. Harris, E. Marinari, E. Martinec, and M. Picco, Phys. Rev. Lett. \textbf{71}, 811 (1993).
		
		\bibitem{behnam}
		B. Tavagh-Mohammadi, M. Masihi, and M. Ganjeh-Ghazvini, Physica A: Statistical Mechanics and its Applications \textbf{460}, 304 (2016).
		
		
		\bibitem{beer}
		T. Beer and I. G. Enting, Math. Comput. Model. \textbf{13}, 77 (1990).
		
		\bibitem{solomon}
		S. Solomon, G. Weisbuch, L. de Arcangelis, N. Jan, and D. Stauffer, Physica A \textbf{277}, 239 (2000).
		
		
		\bibitem{goldenberg}
		J. Goldenberg, B. Libai, S. Solomon, N. Jan, and D. Stauffer, Physica A \textbf{284}, 335 (2000).
		
		
		\bibitem{ball} 
		Z. Ball, H. M. Phillips, D. L. Callahan, and R. Sauerbrey, Phys. Rev. Lett. \textbf{73}, 2099 (1994).
		
		\bibitem{coniglio}
		A. Coniglio, H. E. Stanley, and W. Klein, Phys. Rev. Lett. \textbf{42}, 518 (1979).
		
		\bibitem{henley}
		C. L. Henley, Phys. Rev. Lett. \textbf{71}, 2741 (1993).
		
		\bibitem{cardy}
		J. L. Cardy and P. Grassberger, J. Phys. A \textbf{18} L267 (1985).
		
		\bibitem{moore}
		C. Moore and M. E. J. Newman, Phys. Rev. E \textbf{61}, 5678 (2000).
		
		\bibitem{yara}
		Y. Kanai, K. Abe, and Y. Sekic, Physica A \textbf{427}, 226 (2015).
		
		\bibitem{davis}
		S. Davis, P. Trapman, H. Leirs, M. Begon, and J. A. P. Heesterbeek, Nature \textbf{454}, 634 (2008).
		
		\bibitem{cohen}
		R. Cohen, K. Erez, D. ben-Avraham, and S. Havlin, Phys. Rev. Lett. \textbf{85}, 4626 (2000).
		
		\bibitem{moreira}
		A. A. Moreira, J. S. Andrade Jr., H.J. Herrmann, and J. O. Indekeu, Phys. Rev. Lett. \textbf{102}, 018701 (2009).
		
		\bibitem{de_lima}
		B.N.B. de Lima, R. Sanchis, D.C. dos Santos, V. Sidoravicius, and R. Teodoro, Stoch. Process. Appl. \textbf{130}, 5492 (2020).
		
		\bibitem{grimmett}
		G. Grimmett, \emph{Percolation}, 2nd ed. (Springer-Verlag, Berlin, 1999).
		
		
		\bibitem{gaunt}
		D. S. Gaunt, A. J. Guttmann, and S. G. Whittington, J. Phys. A \textbf{12}, 75 (1979).
		
		\bibitem{kertesz}
		J. Kertesz, B. K. Chakrabarti, J. A. M. S. Duarte, J. Phys. A \textbf{15}, L13 (1982).
		
		
		\bibitem{furlan}
		A. P. Furlan, D. C. dos Santos, R. M. Ziff, and R. Dickman, To appear in Physical Review Research.		
		\bibitem{soteros}
		C. E. Soteros, K. S. S. Narayanan, K. De'Bell, and S. G. Whittington,  Phys. Rev. E, \textbf{53}, 4745 (1996).
		
		\bibitem{wilkinson1}
		M. K. Wilkinson,  J. Phys. A \textbf{19}, 3431 (1986).
		
		\bibitem{holroyd}
		A. E. Holroyd and Z. Li, arXiv:1510.03943v2 (2016).
		
		\bibitem{garet}
		O. Garet, R. Marchand, and I. Marcovici, ALEA, Lat. Am. J. Probab. Math. Stat. \textbf{15}, 279 (2018). 	
			
		\bibitem{harris}
		T. E. Harris, Math. Proc. Cambridge \textbf{56}, 13	(1960).
		
		\bibitem{kesten}
		H. Kesten, Commun. Math. Phys. \textbf{74}, 41 (1980).
		
		\bibitem{ziff01}
		M. E. J. Newman and R. M. Ziff, Phys. Rev. E \textbf{64}, 016706 (2001).
		
		
		\bibitem{leath}
		P. L. Leath, Phys. Rev. Lett. \textbf{36}, 921 (1976).
		
		\bibitem{lenormand}
		R. Lenormand and S. Bories, C. R. Hebd. Seances Acad. Sci. B \textbf{291}, 279 (1980).
		
		\bibitem{chandler}
		R. Chandler, J. Koplik, K. Lerman, and J.F. Willemsen, J. Fluid Mech. \textbf{119}, 249 (1982).
		
		\bibitem{wilkinson2}
		D. Wilkinson and J.F. Willemsen, J. Phys. A \textbf{16}, 3365 (1983). 
		
		
		
		\bibitem{grimmett_prob}
		G. Grimmett and D. Welsh, \textit{Probability: an introduction}, 2nd ed. (Oxford University Press., 2014).
		
		\bibitem {choi}
		J. O. Choi and U. Yu, J. Comp. Phys. \textbf{386}, 1 (2019).
		
		
		\bibitem{stauffer94}
		D. Stauffer and A. Aharony, {\it Introduction To Percolation Theory}, 2nd ed.
		(Taylor $\&$ Francis, London, 1994).
		
		\bibitem{lorenz_d3_1}
		C. D. Lorenz and R. M. Ziff, Phys. Rev. E \textbf{57}, 230 (1998). 
		
		
		\bibitem{dammer_d3_2}
		S. M. Dammer and H. Hinrichsen,  J. Stat. Mech.  \textbf{2004}, P07011 (2004).
		
		\bibitem{wang_d3_3}
		J. Wang, Z. Zhou, W. Zhang, T. M. Garoni, and Y. Deng,  Phys. Rev. E \textbf{87}, 052107 (2013).
		
		\bibitem{mertens_d4_1}
		S. Mertens and C. Moore, Phys. Rev. E \textbf{98}, 022120 (2018).
		
		\bibitem{xun_d4_2}
		Z. Xun and R. M. Ziff, Phys. Rev. Res. \textbf{2}, 013067 (2020).
		
		\bibitem{xu}
		X. Xu, J. Wang, J. P. Lv, and Y. Deng, Front. Phys. \textbf{9}, 113 (2014).
		
		\bibitem{koza}
		Z. Koza, and J. Poła, J. Stat. Mech. \textbf{2016}, 103206 (2016).
		
		\bibitem{zhang}
		Z. Zhang, P. Hou, S. Fang, H. Hu, and Y. Deng,  arXiv:2004.11289 (2020).
		
		
		
		
		
		
	\end{thebibliography}
\end{document}